\documentclass[twocolumn,amsmath,amssymb,superscriptaddress,longbibliography]{revtex4-1}

\newcommand{\IMSS}{Muon Science Laboratory, Institute of Materials Structure Science, High Energy Accelerator Research Organization (KEK-IMSS), Tsukuba, Ibaraki 305-0801, Japan}
\newcommand{\Sokendai}{The Graduate University for Advanced Studies (Sokendai), Tsukuba, Ibaraki 305-0801, Japan}
\newcommand{\JAEA}{Advanced Science Research Center, Japan Atomic Energy Agency, Tokai, Ibaraki 319-1195, Japan}

\usepackage{multirow}
\usepackage{graphicx}
\usepackage[dvipdfmx]{color}
\usepackage{dcolumn}

\usepackage{txfonts}
\usepackage{bm}
\usepackage{ulem}

\usepackage[utf8]{inputenc}
\usepackage[T1]{fontenc}
\usepackage{mathptmx}
\usepackage{etoolbox}
\usepackage[hypertex,colorlinks=true,linkcolor=black,citecolor=blue,filecolor=blue,urlcolor=blue,setpagesize=false,nesting=true]{hyperref}

\makeatletter
\def\@email#1#2{%
 \endgroup
 \patchcmd{\titleblock@produce}
  {\frontmatter@RRAPformat}
  {\frontmatter@RRAPformat{\produce@RRAP{*#1\href{mailto:#2}{#2}}}\frontmatter@RRAPformat}
  {}{}
}%
\makeatother

\begin{document}
\title{Local electronic structure and dynamics of hydrogen in $\text{CeO}_2$}

\author{A.~Koda}\affiliation{\IMSS}\affiliation{\Sokendai}
\author{T.~U.~Ito}\affiliation{\JAEA}
\author{M.~Hiraishi}\affiliation{\IMSS}
\author{H.~Okabe}\affiliation{\IMSS}
\author{R.~Kadono}\thanks{email: ryosuke.kadono@kek.jp}\affiliation{\IMSS}


\begin{abstract}

The local electronic states of muon (Mu) as an isotope of hydrogen (H) in high-quality single-crystalline ceria (CeO$_2$) are investigated using muon spin rotation/relaxation ($\mu\text{SR}$) and first-principles density functional theory (DFT) calculations. Upon positive muon implantation, both paramagnetic (Mu$^0$) and diamagnetic (Mu$^*$) states are observed below $\approx$10 K. Magnetic field dependence of the Mu$^0$ signals combined with DFT results provides evidence for the formation of a polaron state, consisting of Mu bonded to a ligand oxygen and a $4f$ electron localized on a nearby Ce site. The crystal-orientation dependence of the Mu$^0$ signal suggests strong anisotropy of the $4f$ electron due to the spin-orbit coupling with lifted degeneracy. Furthermore, the Mu$^0$ state exhibits transition to the Mu$^*$ state that corresponds to another Mu$^0$ state (exhibiting a diamagnetic response due to fast spin/charge fluctuations) at temperatures above $\approx$10 K, before disappearing above $\approx$30 K. These findings suggest rapid diffusive motion of the $4f$ electrons and/or Mu (as well as H) at higher temperatures. 
\end{abstract}
\maketitle

Ceria (CeO$_2$) is known to exhibit excellent oxygen storage capacity and high ionic conductivity, and currently used in a variety of applications such as a promoter in automotive three-way catalysts, an electrolyte in solid oxide fuel cells, and an active element in oxygen sensors \cite{Montini:16,Skorodumova:02}. In these applications, the surface of ceria, particularly the oxygen vacancy sites created by reduction, exhibits high affinity for water molecules: they supply hydroxyl groups through the dissociative adsorption of water molecules (H$_2$O $\rightarrow$ H$^+$ + OH$^-$), thereby promoting the oxidation reaction \cite{Montini:16,Bunluesin:98}. This strongly suggests that the behavior of H on its surface and within the bulk material of ceria is a critically important factor governing its ionic/electrical conductivity and catalytic activity.

From the viewpoint of defect physics and chemistry, ceria is an oxide semiconductor with a wide band gap of approximately 3 eV \cite{Wuilloud:84} (close to that of TiO$_2$), making it an interesting research subject within the general context of the local electronic structure of H as an amphoteric impurity \cite{Kilic:02,Walle:03,Peacock:03,Xiong:07}. Recent calculations based on first-principles density functional theory (DFT) suggest that interstitial H in ceria behaves as a positive-$U$ center, where the most stable state involves an OH bond with ligand oxygen, creating two impurity levels within the band gap, $E^{+/0}$ and $E^{0/-}$  \cite{Zacherle:13,Hoang:25}. This indicates that H exhibits successive transition of charge states from H$^+$ to H$^0$, and then to H$^-$ with increasing Fermi energy, which is in marked contrast with other oxides where the interstitial H behaves as a negative-$U$ center accompanying double charge transition level ($E^{+/-}$) in the thermal equilibrium state \cite{Kilic:02,Walle:03,Peacock:03,Xiong:07}. Particularly in strongly correlated electron systems like ceria, where the unoccupied (conduction) band is primarily composed of $f$-electron orbitals, the question of whether the electron supplied by H is in a ``shallow state'' excited into the conduction band \cite{Walle:03b} or localized as a polaron remains an unresolved issue in both theory and experiment.  However, experimental techniques capable of directly observing the electronic states of trace amounts of H are limited, and to date, there is very little experimental information available at the atomic scale.

Muon spin rotation ($\mu$SR) is one such technique, where studies on the local electronic structure of implanted positive muon ($\mu^+$) as a light pseudo-isotope of the proton (about 1/9th of the proton mass) in matter can provide information on the local electronic states of H present at extremely low concentrations within bulk materials (hereafter, the muon as a hydrogen isotope is denoted by the element symbol ``Mu''). Pioneering $\mu$SR studies on ceria also exist, with experiments using powder samples reporting the presence of a paramagnetic state (Mu$^0$, a bound state of $\mu^+$ and an unpaired electron $e^-$) at low temperatures \cite{Cox:06b}. However, the powder-averaged signal did not yield information about the precise Mu site(s) nor the anisotropy of its hyperfine parameters determined by the local electronic structure.

\begin{figure}[t]
  \centering
	\includegraphics[width=0.9\linewidth,clip]{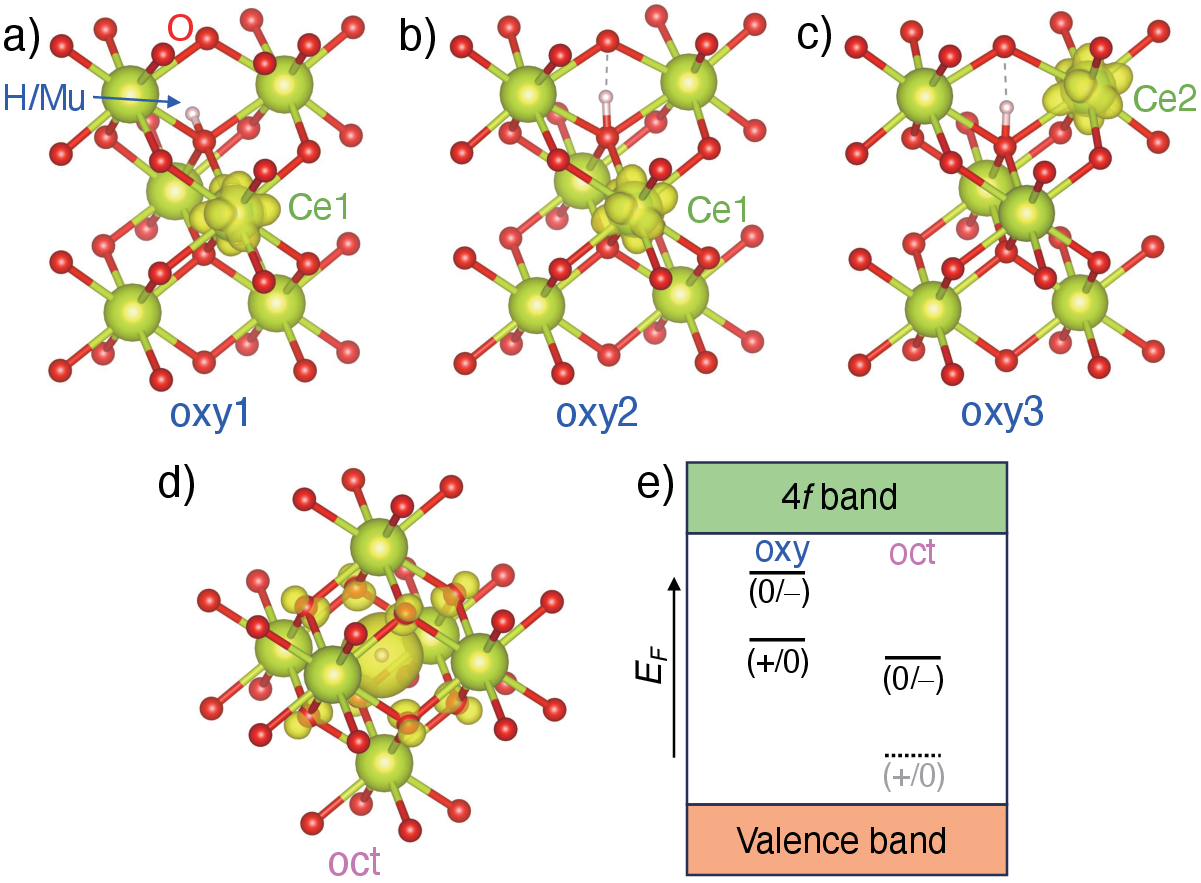}
	\caption{
	(a)--(d) Electronic structures of interstitial H in CeO$_2$ predicted by DFT calculations: H is bonded to O with an unpaired $f$ electron located at (a), (b) Ce1 site, or (c) Ce2 site. The H-O-Ce1 angle in (a) is $\approx$180$^\circ$ (oxy1). The O-H-O angle (connected by dashed lines) is (b) $\approx$168$^\circ$ (oxy2) and (c) $\approx$175$^\circ$ (oxy3). (d) Atom-like H is located at the Ce-cornered octahedral (oct) site.  (e) Schematics of the electronic levels associated with H in the energy band diagram (after Ref.[\onlinecite{Zacherle:13}]), where the $E^{(+/0)}$ level for the oct configuration is inaccessible due to the instability of the H$^+$ state. }
	\label{dft}
\end{figure}

In this study, detailed $\mu$SR measurements using high-quality single crystals and DFT calculations were performed. Our experiment revealed coexistence of paramagnetic (Mu$^0$) and diamagnetic (Mu$^*$) states at low temperatures, which is in line with the recently developed ``ambipolarity model'' \cite{Hiraishi:22,Kadono:24a}. Furthermore, by comparing the magnetic field dependence and crystal orientation dependence of the signal from the Mu$^0$ state with DFT calculation results, we found evidence that Mu$^0$ is a polaron state consisting of Mu bonded to ligand oxygen (analogous to the OH bond) and an electron localized at the neighboring Ce site (attributed to Mu$^0_{\rm oxy2}$ in Fig.~\ref{dft}(b)).  The temperature dependence of the relative yields and spin relaxation of the two Mu states suggests that Mu$^*$ is another atom-like Mu$^0$ state (attributed to Mu$^*_{\rm oct}$ in Fig.~\ref{dft}(d)) which is in the motional narrowing regime (denoted by ``$*$''). Based on these results, we infer the electronic structure and diffusion kinetics of H in ceria.

The samples used in the $\mu$SR experiment were two pieces of single crystalline ceria (7$\times$7$\times$0.5 mm$^3$) with $\langle100\rangle$ orientation (provided by SurfaceNet GmbH via Crystal Base, Co. Ltd.).
Conventional $\mu$SR measurements were performed under zero field (ZF), longitudinal field (LF, the external field $B_{\rm LF}$ parallel with the initial muon polarization $P_z(0)$) and weak transverse field (TF, $B_{\rm TF}\perp P_z(0)$)  using the S1 instrument (ARTEMIS)~\cite{Kojima:14} of the Materials and Life science Experiment Facility (MLF), J-PARC. A nearly 100\% spin-polarized $\mu^+$ beam (with a kinetic energy of $\sim$4~MeV and mean pulse width $T_{\rm w}\approx0.1$ $\mu$s, corresponding frequency resolution $f_{\rm N}=1/2T_{\rm w}\approx5$ MHz) was stopped in the sample mounted on the cold finger (made of 99.99\% silver) attached to the He gas-flow cryostat for controlling sample temperature. The single muon pulse beam measurements were performed for the crystal orientation $\langle100\rangle\parallel B_{\rm LF}$.  Additional measurements regarding crystal orientation dependence were performed using a double-pulse beam (pulse interval $\approx0.6$ $\mu$s) under the conditions $\langle110\rangle\parallel B_{\rm LF}$ and $\langle111\rangle\parallel B_{\rm LF}$, which were achieved by tilting and/or rotating the sample holder relative to the LF direction. The time-dependent $\mu^+$ spin polarization, which reflects the local magnetic field at the Mu site, was monitored via the forward-backward asymmetry [$A(t)$] of $\mu$-$e$ decay positrons. The observed $A(t)$ (the $\mu$SR time spectrum) was analyzed by least-squares ($\chi^2$) curve fit using ``musrfit" code~\cite{musrfit}.
The asymmetry of background signal from muons which missed the sample was estimated using a calibration data obtained by Ho slabs of various cross section to be $A_{\rm bg}\approx0.05$ for $\langle100\rangle\parallel B_{\rm LF}$, and $A_{\rm bg}\approx0.06$ for the other orientations.

To estimate the atomic configuration of H-related defects in CeO$_2$, we performed structural optimizations of model supercells within the DFT+$U$ framework using the Vienna Ab initio Simulation Package (VASP) \cite{Kresse:96,Kresse:99}. The details of our calculations are found in Supplemental Material (SM) \cite{SM}. The structure of the H$^0$ state obtained from our calculations is shown in Fig.~\ref{dft}(a)-(d). The $4f$ electron in (a)--(c) is localized at the Ce site next to the OH group, thus predicting the polaron state (labeled H$_{{\rm oxy}i}$, $i=1$--3): the estimated axial dipolar hyperfine parameters $\omega_\parallel/2\pi$ are (a) 11.3 MHz, (b) 14.0 MHz, and (c) 47.0 MHz \cite{SM}. Specifically, the $4f$ electron is localized at the Ce1 site in (a) and (b), while it is localized at the Ce2 site in (c). Furthermore, the O-H axis in (a) is nearly parallel to the O-Ce1 axis (i.e., H is at the O-Ce anti-bonding site), whereas the angle formed by the OH group and the neighboring O atom in (b) and (c) is close to 180$^\circ$. From the viewpoint of formation energy $E_{\rm H}$, (a) is the most stable configuration with slightly higher energy for (b) ($\approx$0.0438 eV). However, prior calculations suggest that (b) is the lowest energy state \cite{Zacherle:13,Hoang:25}, indicating that these two states are comparable in $E_{\rm H}$. The configuration (c) is 0.101 eV higher in $E_{\rm H}$ than (a), suggesting an increase in Coulomb energy due to the shortened distance H$^+$-Ce$^{3+}$.  In contrast, the structure of H in (d) (H$_{\rm oct}$) is atom-like (the estimated Fermi-contact hyperfine parameter $\omega_0/2\pi=3471.5$ MHz \cite{SM}), and $E_{\rm H}$ is approximately 2 eV higher than that of H$^0_{\rm oxy}$. Moreover, while the H$_{\rm oct}^0$ and H$_{\rm oct}^-$ states can exist as metastable states, the H$_{\rm oct}^+$ state is unstable as it resides at a saddle point of the configuration potential and undergoes a transition to H$_{\rm oxy}$. 

As has been demonstrated in the ambipolarity model \cite{Kadono:24a}, both H$_{\rm oxy}$ and H$_{\rm oct}$ states can be accessed by implanted Mu despite relatively large $E_{\rm H}$ for the latter. Therefore, if a potential barrier exists for the transition between them, the H$_{\rm oct}$ state could also exist as a metastable state. Furthermore, the Fermi energy dependence of the formation energy, 
\begin{equation}
E^q_{\rm H}(E_F) = E_{\rm tot}[{\rm H}^q] - E_{\rm tot}[\text{--}] + qE_F - n_{\rm H}\mu_{\rm H},
\end{equation}
 (where $E_F$ is the Fermi energy measured from the valence band top, $q = +1,$ 0, $-1$ corresponds to H$^+$, H$^0$, H$^-$, $E_{\rm tot}[X]$ is the total energy with or  without defect $[X={\rm H}^q$ or null$]$, $n_{\rm H}$ is the number density of H, and $\mu_{\rm H}$ is the chemical potential of H) indicate that both H$_{\rm oxy}$ and H$_{\rm oct}$ states behave as positive-$U$ centers \cite{Zacherle:13}.  The thermodynamic charge-transition levels associated with the local H states determined by the intersections of $E^+_{\rm H}$ and $E^0_{\rm H}$ and that of $E^0_{\rm H}$ and $E^-_{\rm H}$ are schematically shown in Fig.~\ref{dft}(e). The $E^{(0/-)}$ level for H$_{\rm oxy}$ is close to the 4$f$ band (considered the conduction band), and H$_{\rm oxy}$ can behave as a shallow donor.  The $E^{(0/-)}$ level for H$_{\rm oct}$ is situated near the midgap, and H$_{\rm oct}$ is prone to a deep acceptor.


\begin{figure}[t]
  \centering
	\includegraphics[width=0.7\linewidth,clip]{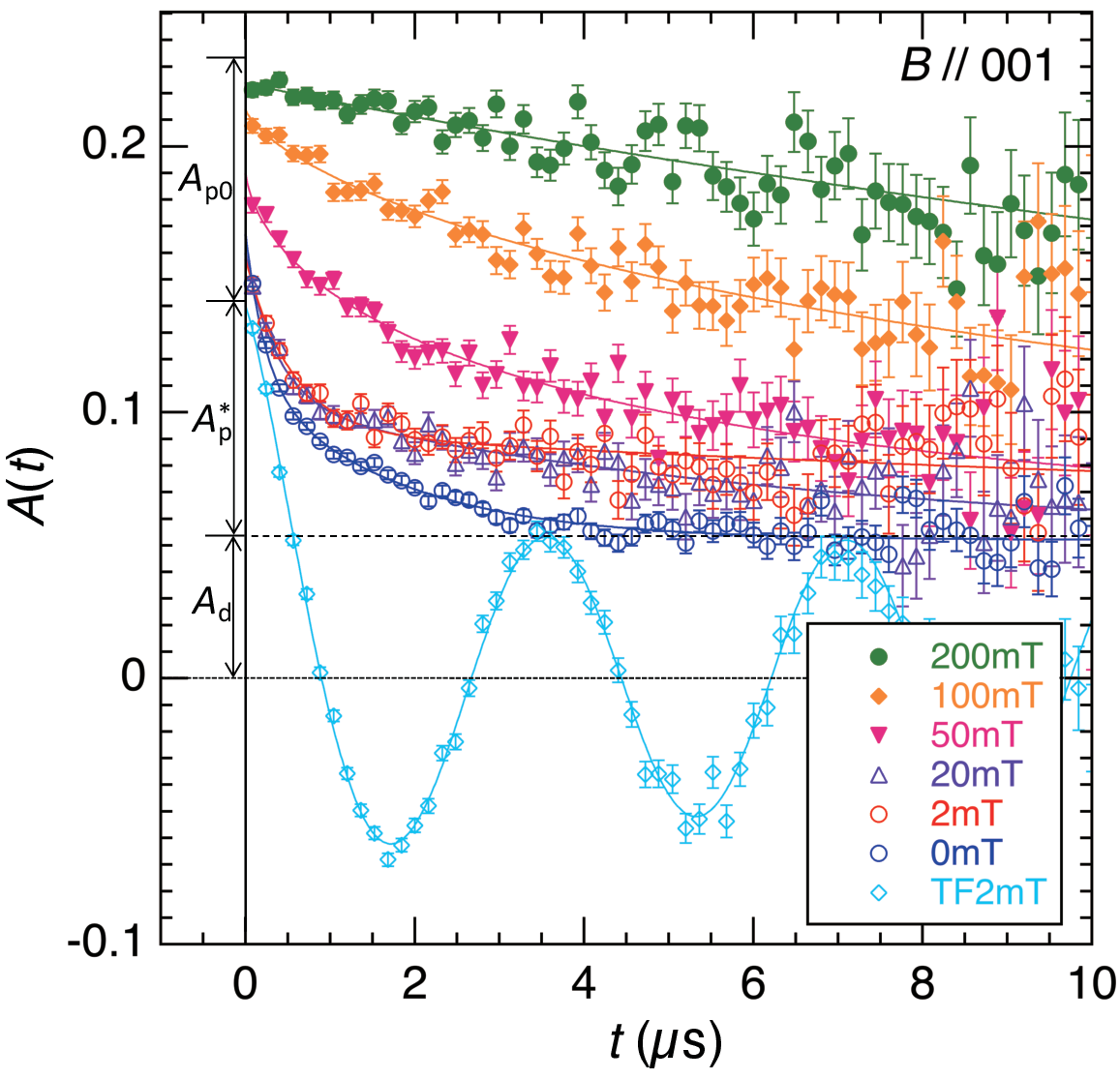}
	\caption{
	$\mu$SR time spectra observed at 8.8~K under various external magnetic fields. The spectrum under TF (= 2 mT) represents signals from the diamagnetic Mu$^*$ state with the partial asymmetry $A^*_{\rm p}$ and the non-relaxing diamagnetic Mu signal(s) with $A_{\rm d}$ (including background), whereas those under LF (= 0 to 300 mT) involve the signal from Mu$^0$ state whose partial asymmetry $A_{\rm p}(B)$ varies with LF (where $A_{\rm p0}$ is the maximum value for $B\rightarrow\infty)$.}
	\label{tspec}
\end{figure}

Figure \ref{tspec} shows the $\mu$SR time spectra measured at 8.8 K with various LFs applied parallel to the $\langle100\rangle$ axis, which are overlayed with the TF-$\mu$SR spectrum ($B_{\rm TF}=2$ mT $\perp\langle100\rangle$). The envelope of the TF-$\mu$SR spectrum consists of an exponentially decaying component and a non-relaxing component respectively given by the partial asymmetry $A^*_{\rm p}$ and $A_{\rm d}$, where the latter includes the non-relaxing diamagnetic Mu component from the sample and the background ($A_{\rm bg}$). Meanwhile, the magnitude of $A(0)$ increases beyond $A^*_{\rm p}+A_{\rm d}$ as LF increases within the 10$^2$ mT range, which is typical of the signal from a quasi-static Mu$^0$ state. Furthermore, detailed examination of $A(t)$ under LF reveals that these spectra mainly consist of two components which are discerned by the difference in the LF dependence of the partial asymmetry and relaxation rate, namely, (1) a quasi-static Mu$^0$ component whose partial asymmetry $A_{\rm p}(B)$ and relaxation rate $\lambda(B)$ strongly depend on LF, and (2) a diamagnetic Mu$^*$ component corresponding to $A_{\rm p}^*$ exhibiting relatively slow relaxation rate $\lambda^*(B)$ that is only weakly dependent on LF.  The best-fit curves obtained by analyzing these spectra using $\chi^2$ fits with the phenomenological relaxation functions for TF,
\begin{equation}
A(t)=(A^*_{\rm p}e^{-\lambda_{\perp} t}+A_{\rm d})\cos\omega_\mu t,\label{tfs}
\end{equation}
and LF,
\begin{equation}
A(t)=A^*_{\rm p}e^{-\lambda^*(B) t}+A_{\rm p}(B)e^{-\lambda(B) t}+A_{\rm d},\label{lfs}
\end{equation}
are shown as solid lines in Fig.~\ref{tspec}, where $\omega_\mu=\gamma_\mu B_{\rm TF}$ is the Larmor precession frequency of $\mu^+$ with $\gamma_\mu$ ($=2\pi\times135.53$ MHz/T) being the muon gyromagnetic ratio, $\lambda_{\perp}$ is the transverse spin relaxation rate. 
Note that the quasi-static Mu$^0$ exhibits an effective gyromagnetic ratio of $\frac{1}{2}(\gamma_e+\gamma_\mu) = 2\pi\times14.08$ MHz/mT at low $B_{\rm TF}$ (with $\gamma_e=2\pi\times28024.2$ MHz/T being the electron gyromagnetic ratio), comprising a signal with a frequency of $\approx$28 MHz at 2 mT. However, this far exceeds the resolution $f_{\rm N}\approx5$ MHz in the present measurements, leading to $A_{\rm p} = 0$ in Eq.~(\ref{tfs}).

\begin{figure}[t]
  \centering
	\includegraphics[width=0.65\linewidth,clip]{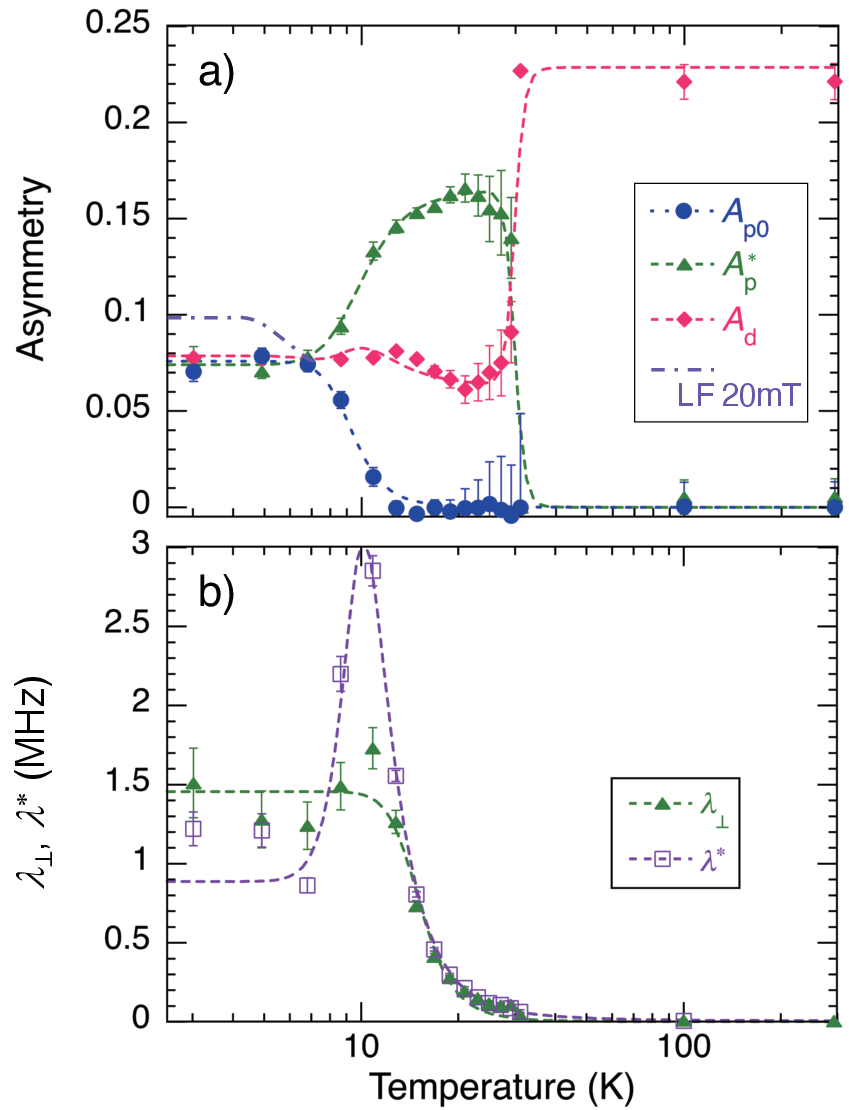}
	\caption{
	Temperature dependence of (a) the partial asymmetry $A_{\rm p0}$ (quasi-static Mu$^0$), $A_{\rm p}^*$ (Mu$^*$), and diamagnetic Mu, where the dot-dashed curve shows the behavior of $A_{\rm p}$ at LF = 20 mT. (b) The relaxation rate of the $A_{\rm p}^*$ component under TF = 2 mT ($\lambda_\perp$) and LF = 20 mT ($\lambda^*$) versus temperature. (See the main text for the dashed curves.) }
	\label{tfasy}
\end{figure}

While our primary interest lies in the LF dependence of the signal from Mu$^0$ and the electronic structure it suggests, we first examine the relative stability of the Mu$^0$ and Mu$^*$ states from the temperature dependence of their partial asymmetry deduced from TF-$\mu$SR spectra. As shown in Fig.~\ref{tfasy}(a), $A_{\rm p}^*$ increases sharply with temperature above $\approx$7 K from a constant value, exhibits a broad peak between 10 and 30 K, and then abruptly disappears above $\approx$30 K. In contrast, $A_{\rm d}$ remains nearly constant below 30 K, except for a slight decrease between 10--30 K, and then increases sharply above 30 K in place of $A_{\rm p}^*$. These behaviors respectively match the temperature dependence for the ``Mu$^0$ (fast relaxing)'' and ``Mu$^+$ (slow relaxing)'' components reported in the literature \cite{Cox:06b}. Since $A_{\rm d}$ above 30 K exhibits a value corresponding to 100\% muon polarization ($\approx$0.23), subtracting $A_{\rm p}^*+A_{\rm d}$ from this value yields $A_{\rm p0}$. This assignment was confirmed in a similar temperature scan performed at LF = 20 mT, where a partial recovery in $A_{\rm p}$ (i.e., LF-decoupling) was observed below $\approx$7 K (see Fig.~\ref{tfasy}(a)). Moreover, $\lambda^*(20\:{\rm mT})$ in Fig.~\ref{tfasy}(b) exhibits a prominent peak around 10 K, suggesting the sharp increase in the spin fluctuation rate across this temperature (with the peak corresponding to the ``$T_1$-minimum'' \cite{Bloembergen:48}).  These observations suggest that the quasi-static Mu$^0$ state exists only below $\approx$10 K which undergoes thermally activated transition to the Mu$^*$ state, and that it reaches the diamagnetic Mu state above $\approx$30 K.

\begin{figure}[t]
  \centering
	\includegraphics[width=0.9\linewidth,clip]{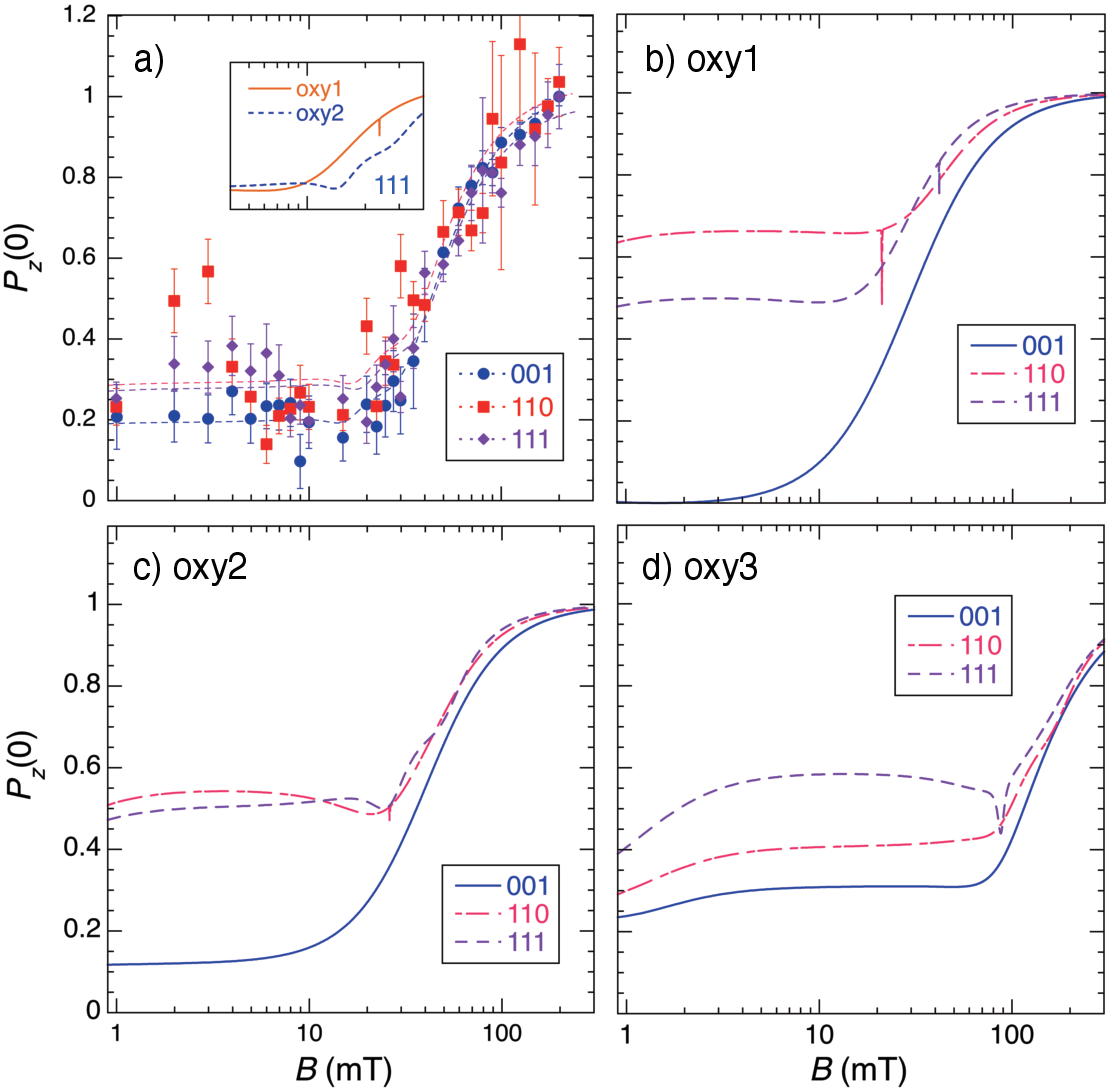}
	\caption{
	(a) LF dependence of the initial polarization $P_z(0)$ observed for LF parallel with the crystal axis $\langle001\rangle$, $\langle110\rangle$, and $\langle111\rangle$ (see text for the dashed curves). The calculated LF dependence of $P_z(0)$ for the corresponding three directions assuming the local structures shown in Fig.~\ref{dft}(a)-(c) with a free $1\mu_B$ spin placed at each Ce site. Inset: comparison of $P_z(0)$ with $B\parallel\langle111\rangle$ between oxy1 and oxy2 over a field range of 4--40 mT around the inflection.}
	\label{LFdcp}
\end{figure}

Next, Fig.~\ref{LFdcp}(a) shows the LF and crystal orientation dependence of the initial polarization $P_z(0) = A_{\rm p}(B)/A_{\rm p0}$ exhibited by the quasi-static Mu$^0$ state. Here, the data for $B\parallel\langle001\rangle$ axis were obtained at 8.8 K, while the data for the other two directions were obtained at 5 K; the relatively large scattering of data points at lower fields in the latter is attributed to the greater ambiguity in determining $A(t)$ near $t=0$ with faster relaxation using a double-pulse beam. 
The first notable feature is that $P_z(0)$ increases sharply with increasing magnetic field for all orientations, exhibiting a steep inflection point around 20--30 mT. Furthermore, it is also noteworthy that the overall LF dependence shows almost no difference among the three orientations.

Let us compare these results with the LF dependence of $P_z(0)$ in Figs.~\ref{LFdcp}(b)--(d) which are predicted for the structures shown in Figs.~\ref{dft}(a)--(c). The calculated curves were derived by numerically solving the spin Hamiltonian for a quantum system of two $S = 1/2$ particles (comprising the muon and a localized electron at each Ce site) where the electron $g$-factor was fixed at 2 and an axial dipolar hyperfine coupling was assumed \cite{Ito:19}.  It can be seen that none of the curve patterns agree with the experimental results. Specifically, for Mu$^0_{\rm oxy1}$ and Mu$^0_{\rm oxy2}$, the prediction of a significant difference between the $\langle001\rangle$ direction and the other two directions does not match the behavior of data at low magnetic fields. In particular for M$^0_{\rm oxy3}$, the relatively large difference in $P_z(0)$  between $\langle110\rangle$ and $\langle111\rangle$, and the significant shift of the inflection point toward the higher field at $\approx$100 mT, do not agree with the experiment; one needs to assume a significantly reduced magnetic moment of $\approx0.35\mu_B$ at the Ce2 site to be consistent with experimental results. Therefore, we hereafter exclude this possibility from consideration. 

The gross mismatch in crystal orientation dependence is presumed to stem from approximating localized $4f$ electron as a free spin, while it must be subject to strong spin-orbit interaction. This naturally leads us to assume that the degeneracy of the $4f$ electron ground state is lifted for some reason (see the following discussion) to induce $g$-tensor anisotropy. In this case, the LF dependence of $P_z(0)$ is fixed to one specific crystal orientation determined by the electron anisotropy regardless of the LF direction. In reexamining the $P_z(0)$ of Mu$^0_{\rm oxy1}$ and Mu$^0_{\rm oxy2}$, the LF dependence for the $\langle111\rangle$ direction of Mu$^0_{\rm oxy2}$ exhibits better agreement with the experimental results regarding the characteristic inflection at 20--30 mT (see inset of Fig.~\ref{LFdcp}(a)). The dashed curves in Fig.~\ref{LFdcp}(a) show the result of curve fits without considering the downward shift of $P_z(0)$ at lower fields (see below), yielding $\omega_\parallel/2\pi=13.9(7)$ MHz, 12.4(6) MHz, and 13.9(8) MHz for $B\parallel\langle001\rangle$, $\langle110\rangle$, and $\langle111\rangle$, respectively.  Thus, Mu$^0_{\rm oxy2}$ becomes a candidate for the local structure of Mu$^0$ with the $\langle111\rangle$ direction for the anisotropy axis of the $4f$ electron. 

\begin{figure}[t]
  \centering
	\includegraphics[width=0.6\linewidth,clip]{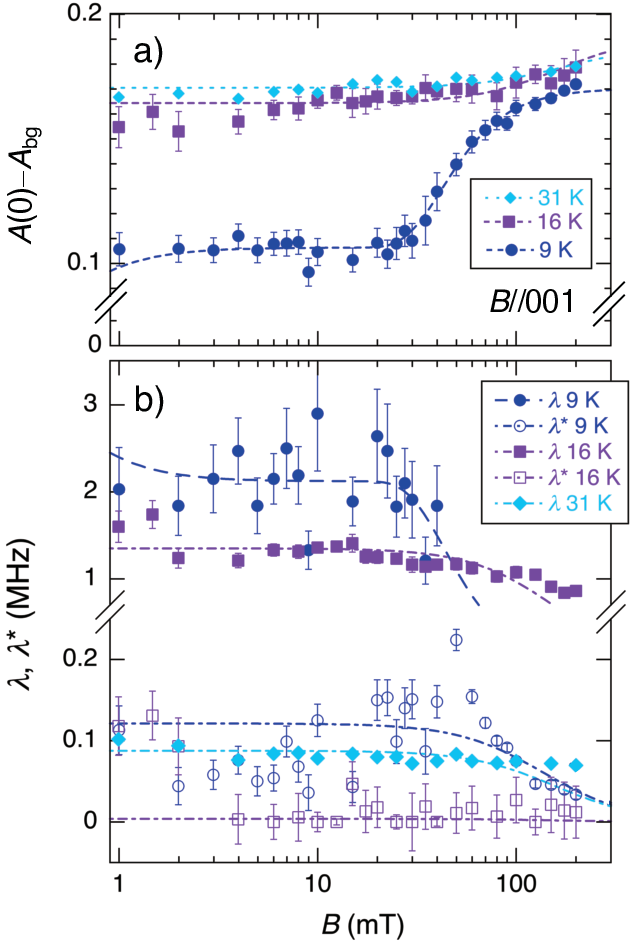}
	\caption{
	 LF dependence of (a) the initial asymmetry $A(0)$ and (b) relaxation rate observed for LF parallel with the crystal axis $\langle001\rangle$. Dashed/dot-dashed curves are results of curve fits assuming the conversion process from quasi-static Mu$^0_{\rm oxy2}$ to dynamical Mu$^*_{\rm oct}$ states described by Eqs.~(\ref{conv}) and (\ref{rlxb}) (see text for detail). }
	\label{LFasyrlx}
\end{figure}

The remaining issue is the discrepancy in $P_z(0)$ below $\approx$20 mT where the data are significantly smaller than the  $\langle111\rangle$ curve in Fig,~\ref{LFdcp}(c) that predicts $P_z(0)\approx0.5$. A possible scenario for this additional depolarization is the conversion reaction from the quasi-static Mu$^0$  to the diamagnetic Mu$^*$ state. This is also inferred from the increase of $A_{\rm p}^*$ in place of $A_{\rm p0}$ and associated enhancement of spin relaxation rate observed above 7 K in Fig.~\ref{tfasy}. (The key point of this scenario is that it describes a transition from the initial state where depolarization occurs due to spin-singlet Mu$^0$ formation to the final state where it does not.) In this case, assuming $P^{\rm i}_z(0)$ for $B\parallel\langle111\rangle$ is the polarization for the initial quasi-static Mu$^0$ state which is Mu$^0_{\rm oxy2}$-like, the final state polarization is given by 
\begin{eqnarray}
P_z^{\rm f}(0) &=& \frac{\kappa}{\lambda(B) + \kappa},\label{conv}\\
\lambda(B) &\approx& \frac{\nu(\omega_\parallel^{\rm i})^2}{2[\nu^2+(\omega_\parallel^{\rm i})^2]}[1 - P^{\rm i}_z(0)], \label{rlxb}
\end{eqnarray} 
where $\nu$ is the fluctuation frequency of the hyperfine parameter $\omega_\parallel^{\rm i}$ for the initial state, and $\kappa$ is the conversion reaction rate to the final state \cite{Kadono:03}. The results of curve fitting using Eq.~(\ref{conv}) for the case of $\nu\ll\omega_\parallel^{\rm i}$ are shown by the dashed line in Fig.~\ref{LFasyrlx}(a), where $A(0)-A_{\rm bg}=A_{\rm p}^*+A_{\rm p0}P^{\rm f}_z(0)$ (see Figs.~S1(a) and (c) in SM \cite{SM} for other orientations). The values of each parameter obtained from this analysis are summarized in the Table \ref{tab1}, where the relatively small $\kappa$ and $\nu$ are consistent with the presumed quasi-static state.  The behavior of $P_z^{\rm f}(0)$ in the low-LF limit is dominated by the ratio between $\kappa$ and $\nu$, and the overall LF dependence is approximately proportional to that of $P^{\rm i}_z(0)$.  The fit curve exhibits excellent agreement with the data with reasonable parameter values, supporting this depolarization scenario with Mu$^0_{\rm oxy2}$ with a slightly reduced hyperfine parameter $\omega_\parallel^{\rm i}$ ($<\omega_\parallel$) as the initial state.


\begin{table}[b]
\begin{tabular}{ccccc}
\hline\hline
 & $\kappa$ (MHz) & $\nu$ (MHz) & $\omega_\parallel^{\rm i}/2\pi$ (MHz) & Temp. (K)\\
 \hline
 $\langle001\rangle$ & 0.059(5) & 0.997(3) & 8.1(1.5) & 8.8 \\
 $\langle110\rangle$ & 0.101(15) & 0.994(11) & 9.4(1.3)& 5 \\
 $\langle111\rangle$ & 0.096(8) & 1.003(7) & 10.0(8)& 5 \\
\hline\hline
\end{tabular}
\caption{The parameters obtained by curve fitting of $P_z(0)$ versus $B_{\rm LF}$ in Fig.~\ref{LFasyrlx}(a) using Eq.~(\ref{conv}). }\label{tab1}
\end{table}

The above model can be further confirmed by the detailed LF dependence of the spin relaxation rate at 8.8 K shown in Fig.~\ref{LFasyrlx}(b) for $B\parallel\langle001\rangle$ (see SM \cite{SM} for the other two directions).  Here, $\lambda(B)$, showing a relatively large relaxation rate at lower fields, decreases sharply above $\approx$20 mT which is well described by Eq.~(\ref{rlxb}) (thus becoming indistinguishable from $\lambda^*(B)$ above around 50 mT). In contrast, $\lambda^*(B)$ remains nearly constant up to $\approx$100 mT and decreases slightly at higher fields. This suggests that the Mu$^*$ state actually corresponds to a paramagnetic state and that the hyperfine parameter $\omega_0$ causing the relaxation is significantly larger than $\omega_\parallel$. Furthermore, the weak LF dependence of $A_{\rm p}^*$ indicates that this component is subject to the motional narrowing \cite{Kadono:25}. Therefore, the corresponding fluctuation frequency $\nu^*$ is presumed to be greater than $\omega_0$. In this case, the LF dependence of $\lambda^*(B)$ is expected to be described by Eq.~(\ref{rlxb}) with  $\omega_\parallel^{\rm i}$ and $\nu$ replaced by $\omega_0$ and $\nu^*$, respectively (corresponding to that for $\nu^*\gtrsim\omega_0$). The dashed curves in Fig.~\ref{LFasyrlx}(b) represent the result obtained by curve fitting assuming $\omega_0$ is isotropic and setting $P^{\rm i}_z(0) = (1/2 + x^2)/(1 + x^2)$ (where $x=(\gamma_\mu-\gamma_e)B/\omega_0$, which reasonably reproduces the data behavior with $\omega_0/2\pi=3.9(2)$ GHz with $\nu^*$ much greater than $\omega_0$ (see SM \cite{SM} regarding other parameters). The value of $\omega_0$ is consistent with that for the Mu$^0_{\rm oct}$ state predicted by our DFT calculation \cite{SM}. Therefore, we can attribute the final state of this conversion process to Mu$^*_{\rm oct}$. This is also supported by the LF-dependence of $A(0)$ and $\lambda^*(B)$ at 16 and 31 K shown in Fig.~\ref{LFasyrlx}, where the inflection around 20--30 mT characteristic to Mu$^0_{\rm oxy2}$ disappears at these temperatures and the LF-dependence is better described by that for the Mu$^*_{\rm oct}$ state.


Now, let us discuss the origins of (i) the presumed 4$f$ electron anisotropy, (ii) conversion from Mu$^0_{\rm oxy2}$ to the Mu$^*_{\rm oct}$ around 10 K, (iii) the spin relaxation of bound electron in these Mu states, and (iv) the sharp disappearance of the Mu$^*_{\rm oct}$ state above $\approx$30 K. 

Regarding (i), In the cubic crystalline electric field ($O_h$) of the stoichiometric lattice, the Ce$^{3+}(4f^1)$ levels split into a $\Gamma_7$ doublet and a $\Gamma_8$ quartet. While these multiplets are energetically close in the eight-fold coordination of CeO$_2$, experimental evidence from magnetic susceptibility measurements on donor-doped CeO$_2$ consistently suggests that the $\Gamma_8$ quartet constitutes the ground state \cite{Kolodiazhnyi:17}. In addition, the introduction of an interstitial  Mu induces a local symmetry breaking, acting as a Mu-induced crystalline electric field perturbation that lifts the four-fold degeneracy of the $\Gamma_8$ state and promotes mixing with the original $\Gamma_7$ component \cite{Feyerherm:95,Tashma:97,Ito:09}. Consequently, the ground state is reconstructed into a Kramers doublet that can be effectively described as a pseudo-spin $S=1/2$ state with an  anisotropic $g$-tensor. This mechanism qualitatively accounts for the observed crystal orientation dependence of $A_{\rm p}(B)$ and the locking of the magnetic anisotropy axis to specific crystallographic directions.

As a crucial prerequisite for discussing (ii) and (iii), it must be remembered that Ce and O are elements without nuclear magnetic moments. This strongly disfavors the diffusion motion of the Mu$^0$ state itself as the cause of electron spin relaxation due to the fluctuation of nuclear hyperfine fields \cite{Kadono:25}. Therefore, its origin must be sought solely in spin/charge exchange interactions of Mu$^0$ with excess electrons/holes located near the Mu$^0$ state. A well-known origin for such electrons/holes is the electronic excitation of valence band electrons due to kinetic energy deposited upon $\mu^+$ implantation \cite{Hiraishi:22,Kadono:24a}. It is presumed that the Mu$^0$ state immediately after $\mu^+$ implantation is also realized by capturing such ``metastable'' electrons excited to the conduction band \cite{Prokscha:07}. Consequently, the spin relaxation of both Mu$^0_{\rm oxy2}$ and Mu$^*_{\rm oct}$ states can also be explained as spin/charge exchange interactions with the other metastable electrons.

The microscopic cause of the conversion from Mu$^0_{\rm oxy2}$ to Mu$^*_{\rm oct}$ remains unclear at this stage. DFT calculations, including our own, indicate that Mu$^0_{\rm oct}$ has a formation energy $\approx$2 eV higher than Mu$^0_{\rm oxy2}$ \cite{SM}, which apparently disfavors explanation by thermally activated transition. This difficulty may be resolved by considering that the Mu$^0_{\rm oxy2}$ with relatively small $\omega_\parallel^{\rm i}$ corresponds to the ``transition state'' discussed in the Mu generation process in some oxides \cite{Vilao:17,Vilao:18,Vilao:21,Vilao:23}. For example in Al$_2$O$_3$, a transition state trapped in a local potential minimum $\approx$1.3 eV above the final state is predicted to explain the experimental results \cite{Vilao:23}. We also note that Mu$^*_{\rm oct}$ may be such a transition state, given that it disappears above $\approx$30 K.

Regarding the mechanism of (iii), assuming the quasi-Fermi level in the non-equilibrium state lies near the center of the gap in the band structure shown in Fig.~\ref{dft}(e), it is inferred that the $E^{(+/0)}$ level associated with H$_{\rm oxy}$ and the $E^{(0/-)}$ level associated with H$_{\rm oct}$ are involved. However, the difference between $\nu$ and $\nu^*$ may be closely related to the conversion mechanism between the two states, and understanding this remains a future challenge.

Finally, although it is difficult to infer the microscopic mechanism for (iv) solely from these experimental results, analyzing the temperature dependence of $A_{\rm p}^*$ near 30 K using a thermal activation model yields an activation energy $E_{\rm a} \simeq0.07(3)$ eV \cite{SM}. It is interesting whether this is a coincidence that it is comparable to the activation energy ($\simeq0.06$ eV) associated with the diffusion motion of H$^0_{\rm oxy2}$ predicted by DFT calculations \cite{Hoang:25}.

In conclusion, although the electronic excitations associated with $\mu^+$ introduce a certain degree of complexity in the Mu dynamics whose details have not yet been fully elucidated, we demonstrated that Mu, as an isotope of H in CeO$_2$ can adopt two electronic structures: Mu$^0_{\rm oxy2}$ (a polaron state) and Mu$^*_{\rm oct}$ (an atom-like state). DFT calculations suggest that Mu$^0_{\rm oxy2}$ is the possible ground state. Therefore, it is anticipated that the H$^0_{\rm oxy2}$ state is also realized for hydrogen with its valence state depending on the Fermi level.

\begin{acknowledgments}
We would like to thank T.~Matsukawa for helpful discussion. Thanks are also to the MLF staff for their technical support during the $\mu$SR experiment, which was conducted under the support of Inter-University-Research Programs by Institute of Materials Structure Science, KEK (Proposals No. 2013MS01 and 2025MI21). This work was financially supported by the Elements Strategy Initiative to Form Core Research Center for Electron Materials, from the Ministry of Education, Culture, Sports, Science, and Technology of Japan (MEXT) under Grant No.~JPMXP0112101001, and by the MEXT Program: Data Creation and Utilization Type Material Research and Development Project under Grant No.~JPMXP1122683430. M.H. also acknowledges the support of JSPS KAKENHI Grant No.~22K05275 from MEXT.  
\end{acknowledgments}
\vspace{5mm}
\noindent \textbf{Author Contributions}\\
A.K. conceived the study and, along with M.H. and H.O., performed the $\mu$SR experiments. T.U.I. carried out the DFT calculations and simulated the static LF decoupling curves. Data analysis was performed by R.K., who also developed the theoretical model of muon/muonium spin dynamics. R.K. drafted the manuscript with writing contributions from T.U.I. All authors discussed the results and approved the final version of the manuscript.
%
\end{document}